\documentclass[prb,twocolumn]{revtex4-1}
\usepackage{graphicx}
\usepackage{tikz}
\usepackage{mathtools}
\usepackage{amssymb}

\DeclareMathOperator*{\argmax}{arg\,max}

\newcommand{\appropto}{\mathrel{\vcenter{
  \offinterlineskip\halign{\hfil$##$\cr
    \propto\cr\noalign{\kern2pt}\sim\cr\noalign{\kern-2pt}}}}}
    
\begin{document}
\title{Distribution of metastable states of Ising spin glasses}

\author{Stefan Schnabel}
\email{schnabel@itp.uni-leipzig.de}

\author{Wolfhard Janke}
\email{Wolfhard.Janke@itp.uni-leipzig.de}

\affiliation{Institut f\"ur Theoretische Physik, Universit\"at Leipzig, Postfach 100920, 04009 Leipzig, Germany}
\date{\today}

\begin{abstract}

Local minima also known as inherent structures are expected to play an essential role for the behavior of spin glasses. Here, we propose techniques to efficiently sample these configurations in Monte Carlo simulations. For the Sherrington-Kirkpatrick and the three-dimensional Edwards-Anderson model their spectra are determined and compared to analytical results.

\end{abstract}

\maketitle
%%%%%%%%%%%%%%%%%%%%%%%%%%%%%%%%%%%%%%%%%%%%%%%%%%%%%%%%%%%%%%%%%%%%%%%%%%%%%%%%%%%%%%%%%%%%%%%%%%%%%%%%%%%%%%%%%%%%%%%%%%%%%%%%%%%%%%%%%%%%%%%%%%%%%%%%%%%%%%%%%%%%%%%%%%%%%%%%%%%%%%%%%%%%%%%%%%%%%%%%%%
\section{Introduction}

Structurally or magnetically disordered glassy systems show special characteristics like memory effects or replica symmetry breaking as a result of a complex energy landscape. A large number of local minima or metastable states exist and can trap the system  at low temperatures. Therefore, these local energy minima have been of interest for some time. For mean-field spin glasses their distribution as function of energy has been determined as early as 1981.\cite{Bray_Moore} The one-dimensional dimensional system with short-range interactions can also be treated analytically\cite{Li} and some general properties are known for higher-dimensional spin glasses.\cite{Newman_Stein} A completely analytical treatment for the cubic lattice is, however, not available and the numerical investigation of metastable states is even more demanding than standard spin-glass simulations. Previous studies  were restricted to exact enumeration of small systems \cite{Burda_1,Burda_2,Waclaw_Burda} or quenches from random configurations.\cite{BaityJesi_Parisi} In the latter work as well as in studies on structural glasses\cite{Heuer} the local minima are seen in relation to the equilibrium configurations from which they are derived by steepest-descent (greedy algorithm) and are referred to as inherent structures.  Recently we introduced a technique \cite{hoga} for the Edwards-Anderson model that efficiently derives inherent structures from a sequence of spin configurations by means of a dynamic greedy algorithm. In this study we extend this approach and propose a method that samples all metastable states with equal probability. We use this and a more traditional method to measure the distributions of metastable states as a function of energy for the Sherrington-Kirkpatrick (SK)\cite{SK} and the three-dimensional Edwards-Anderson (EA)\cite{EA} model.

The rest of the paper is organized as follows: We discuss the models in section 2 and briefly review the analytical solution of Bray and Moore in section 3. In section 4 we introduce our methods and section 5 contains the results.
% Sample average

%%%%%%%%%%%%%%%%%%%%%%%%%%%%%%%%%%%%%%%%%%%%%%%%%%%%%%%%%%%%%%%%%%%%%%%%%%%%%%%%%%%%%%%%%%%%%%%%%%%%%%%%%%%%%%%%%%%%%%%%%%%%%%%%%%%%%%%%%%%%%%%%%%%%%%%%%%%%%%%%%%%%%%%%%%%%%%%%%%%%%%%%%%%%%%%%%%%%%%%%%%
\section{Model}

We consider the Ising Hamiltonian
\begin{equation}
\mathcal{H}=-\sum\limits_{ \langle ij \rangle } J_{ij}s_is_j,\qquad s_i\in\{-1,1\},
\end{equation}
where the sum runs over all pairs of spins $s_i$ interacting via bonds $J_{ij}$. The latter are randomly chosen according to a Gaussian distribution:
\begin{equation}
P(J_{ij})=\frac1{\sqrt{2\pi J^2}}e^{-J_{ij}^2/2J^2}.
\end{equation}
In case of the SK model every spin interacts with every other while for the 3d EA model spins are placed on the sites of a cubic lattice and only adjacent ones contribute to the energy.
If we consider single-site energies, i.e. the sum of all terms to which an individual spin $s_k$ contributes,
\begin{equation}
e_k=-\sum\limits_{\langle ij\rangle} J_{ij}s_is_j(\delta_{ik}+\delta_{jk})
\end{equation}
one can express the Hamiltonian as 
\begin{equation}
\mathcal{H}=\frac12\sum_ke_k=E
\end{equation}
and the energy change associated with a single spin flip
\begin{equation}
\mathbf{S}=(s_1,\dots,s_N)\rightarrow \mathbf{S'}=(s_1,\dots,s_{k-1},-s_k,s_{k+1},\dots),
\end{equation}
\begin{equation}
e_k\rightarrow e'_k=-e_k
\end{equation}
as functions of it
\begin{equation}
\mathcal{H}(\mathbf{S'})-\mathcal{H}(\mathbf{S})=-2e_k.
\end{equation}
Hence, a metastable state or more precisely a single-flip stable state, i.e., a spin configuration for which every single spin flip causes an increase in energy can be asserted if $e_k<0$ for all $k$. It is the distribution $\Omega(E)$ of these metastable states that we are interested in.

%%%%%%%%%%%%%%%%%%%%%%%%%%%%%%%%%%%%%%%%%%%%%%%%%%%%%%%%%%%%%%%%%%%%%%%%%%%%%%%%%%%%%%%%%%%%%%%%%%%%%%%%%%%%%%%%%%%%%%%%%%%%%%%%%%%%%%%%%%%%%%%%%%%%%%%%%%%%%%%%%%%%%%%%%%%%%%%%%%%%%%%%%%%%%%%%%%%%%%%%%%
\section{Bray and Moore's solution}
\begin{figure}
\begin{center}
\includegraphics[width=.9\columnwidth]{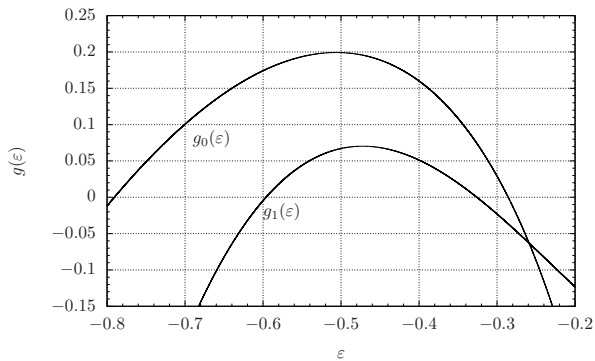}
\caption{\small{\label{fig:BMg} \emph{ The functions $g_0$ and $g_1$.}}}
\end{center}
\end{figure}

In 1981 Bray and Moore \cite{Bray_Moore} derived an analytic expression for the distribution of metastable states for the SK model. They used the dimensionless normalized energy
\begin{equation}
\varepsilon=\frac{E}{NJz^\frac12}
\end{equation}
where $N$ is the total number of spins and $z$ the coordination number of the lattice, i.e., $z=N-1$ for the SK model. For the limit
\begin{equation}
g_0(\varepsilon) \coloneqq \lim\limits_{N \rightarrow \infty}N^{-1}\langle\ln\Omega(\varepsilon)\rangle_J,
\end{equation}
where $\langle\dots\rangle_J$ denotes the disorder average, they obtained
\begin{equation}
g_0(\varepsilon)=\varepsilon^2+2\varepsilon\tau-\ln\left[\sqrt{\pi/2}(-2\varepsilon-\tau)\right],
\end{equation}
where the function $\tau=\tau(\varepsilon)$ is implicitly defined by
\begin{equation}
0=2\varepsilon+\tau+\Phi'(\tau)/\Phi(\tau)
\end{equation}
with
\begin{equation}
\Phi(x)=\frac1{\sqrt{2\pi}}\int\limits_{-\infty}^x e^{-\frac{y^2}2}dy.
\end{equation}
However, they state that this solution is only valid for\\ $\varepsilon > \varepsilon_c \approx -0.672$. Nonetheless, it follows from the position of the maximum of $g_0(\varepsilon)$ shown in Fig.~\ref{fig:BMg} that for large systems the number of metastable states is given by
\begin{equation}
\langle\ln N_S\rangle_J/N=0.199228
\end{equation}
and that the average energy of a local minimum becomes
\begin{equation}
\epsilon_{\rm av}=-0.5061.
\end{equation}

Bray and Moore then proceeded with an expansion in $1/z$ and obtained an approximation for non-mean-field spin glasses:
\begin{equation}
N^{-1}\langle\ln\Omega(\varepsilon)\rangle=g_0(\varepsilon)+z^{-1}g_1(\varepsilon)+O(z^{-2})
\end{equation}
with 
\begin{equation}
g_1(\varepsilon)=-\varepsilon^2(\tau^2-2\varepsilon^2),
\end{equation}
also displayed in Fig.~\ref{fig:BMg}.

%%%%%%%%%%%%%%%%%%%%%%%%%%%%%%%%%%%%%%%%% 
%%%%%%%%%%%%%%%%%%%%%%%%%%%%%%%%%%%%%%%%%

%%%%%%%%%%%%%%%%%%%%%%%%%%%%%%%%%%%%%%%%%%%%%%%%%%%%%%%%%%%%%%%%%%%%%%%%%%%%%%%%%%%%%%%%%%%%%%%%%%%%%%%%%%%%%%%%%%%%%%%%%%%%%%%%%%%%%%%%%%%%%%%%%%%%%%%%%%%%%%%%%%%%%%%%%%%%%%%%%%%%%%%%%%%%%%%%%%%%%%%%%%
\section{Methods}

We apply two methods in order to sample local minima. While the first is a more traditional approach using standard Monte Carlo techniques, the second method is a novel algorithm that has been derived from the dynamic greedy algorithm.\cite{hoga} Its efficiency relies on the specifically local nature of single spin-flips, i.e., low connectivity, and in this study we only apply it to the Edwards-Anderson model.

\subsection{Method I}

This method is a standard Monte Carlo technique which employs flips of single spins and samples in principle all possible states of the spin glass. The ensemble is designed to include all local minima with a sufficiently high probability, such that their distribution can be inferred. Since the goal is to find local minima, i.e., states where all spins have negative energy and which, therefore, are stable against single-spin flips, it is intuitive to use the number of spins with positive energy as a control parameter:
\begin{equation}
    n^{\rm p}(\mathbf{S})=\sum_{i=0}^N\Theta(e_i),
\end{equation}
where $\Theta$ is the Heaviside step function. However, simply minimizing this parameter would yield only local minima around the maximum of the minima distribution, but not in its tails. In order to sample the rare minima at high and at low energy, we also incorporate the Boltzmann weight. In the ensemble $k$ of our simulation a state $\mathbf{S}$ is occupied with a probability
\begin{equation}
    P_k(\mathbf{S}) \propto \omega_k\left( n^{\rm p}(\mathbf{S} ) \right)e^{-\beta_k\mathcal{H}(\mathbf{S})},
\end{equation}
where $\beta_k$ drives the energy of the system similar to the inverse temperature in a canonical ensemble and $\omega_k(m)$ is a weight function that for a given number $m$ equals the inverse sum of the Boltzmann weights of all spin configurations that have $m$ spins with positive energy: 
\begin{equation}
    \omega_k(m) = \left( \sum_\mathbf{S} \delta_{n^{\rm p}(\mathbf{S}),m}\, e^{-\beta_k\mathcal{H}(\mathbf{S})} \right)^{-1},
\end{equation}
where the sum goes over all possible states. The weights $\omega$ causes a flat distribution over $n^{\rm p}(\mathbf{S})$ similar to the weights of a multicanonical simulation\cite{muca1,muca2} or the inverse density of states from the Wang-Landau method\cite{Wang_Landau} leading to a flat histogram over energy. Since $\omega_k$ is a priori not known we determine it before the actual simulation using an iterative procedure.\cite{muca_wght_det} During the simulation a proposed step $\mathbf{S}_{\rm old}\rightarrow\mathbf{S}_{\rm new}$ is accepted with a probability according to the well-known Metropolis criterion:
\begin{equation}
    p^k_{\rm flip}(\mathbf{S}_{\rm old},\mathbf{S}_{\rm new}) = \min\left(1,\frac{P_k(\mathbf{S}_{\rm new})}{P_k(\mathbf{S}_{\rm old})}\right).
\end{equation}

Multiple such ensembles with different $\beta$ are combined via the replica exchange method \cite{parallel_temp1} and two ensembles $k$ and $l$ exchange configurations with the probability
\begin{equation}
    p^{kl}_{\rm exch}(\mathbf{S}_k,\mathbf{S}_l) = \min\left(1,\frac{P_k(\mathbf{S}_l)P_l(\mathbf{S}_k)}{P_k(\mathbf{S}_k)P_l(\mathbf{S}_l)}\right),
\end{equation}
where $\mathbf{S}_k$ is the the configuration belonging to ensemble $k$ before the attempted exchange.

In order to estimate the distribution of local minima we apply the weighted histogram analysis method (WHAM).\cite{wham1,wham2} It is possible to apply this algorithm directly to the various distributions of local minima measured at different $\beta$, however, since the data obtained at low $n^{\rm p}$ is carrying a large statistical error, we decided to determine the reweighting factors using all available data.

We first reweight in order to obtain the unormalized canonical distributions
\begin{equation}
%    \tilde{\Pi}_k(E_i) = \sum_{ \substack{t ,\\ |E_i-\mathcal{H}(\mathbf{S}_{k,t})|<\epsilon/2 } }
    \tilde{\Pi}_k(E_i) = \sum_{ \substack{t ,\\ E_i-\epsilon<E(\mathbf{S}_{k,t})<E_i+\epsilon } }
    \omega_k\left( n^{\rm p}(\mathbf{S}_{k,t} ) \right)^{-1} 
    ,	
\end{equation}
where $\mathbf{S}_{k,t}$ are the configurations generated by the simulation in ensemble $k$ and $2\epsilon$ is the binning width $E_{i+1}=E_i+2\epsilon$, and normalize
\begin{equation}
    \Pi_k(E_i) = \frac{ \tilde{\Pi}_k(E_i) }{ \sum\limits_{j} \tilde{\Pi}_k(E_j) }\approx\frac{g(E_i)e^{-\beta_kE_i}}{z_k}.
\end{equation}
Here, $g(E)$ denotes the density of states and $z_k$ are the partition sums
\begin{equation}
    z_k=\sum_i g(E_i)e^{-\beta_kE_i},
\end{equation}
which can be self-consistently determined by iterating
\begin{equation}
    z_k = \sum\limits_i e^{ -\beta_k E_i } \frac{ \sum\limits_l  \Pi_l(E_i) }{ \sum\limits_l z_l^{-1} e^{ -\beta_l E_i }}.
\end{equation}
One could now calculate the density of states
\begin{equation}
g(E_i)=\frac{ \sum\limits_l  \Pi_l(E_i) }{ \sum\limits_l z_l^{-1} e^{ -\beta_l E_i }}.
\end{equation}
or using the $\beta$-dependent distributions of the local minima,
\begin{equation}
%    \Pi^0_k(E_i) = \frac{\sum\limits_{\substack{t ,\\ |E_i-\mathcal{H}(\mathbf{S}_{k,t})|<\epsilon/2 } } \delta_{n^{\rm p}(\mathbf{S}_{k,t}),0} \omega_k\left( 0 ) \right)^{-1} }{ \sum\limits_{j} \tilde{\Pi}_k(E_j) },
    \Pi^0_k(E_i) = \frac{\sum\limits_{ \substack{t ,\\ E_i-\epsilon<E(\mathbf{S}_{k,t})<E_i+\epsilon } } \delta_{n^{\rm p}(\mathbf{S}_{k,t}),0} \omega_k\left( 0 ) \right)^{-1} }{ \sum\limits_{j} \tilde{\Pi}_k(E_j) },
\end{equation}
derive the overall distribution of local minima
\begin{equation}
\Omega(E_i)=\frac{ \sum\limits_l  \Pi^0_l(E_i) }{ \sum\limits_l z_l^{-1} e^{ -\beta_l E_i }}.
\end{equation}

\subsection{Method II}

\paragraph{Basic concept}
The aim of this method is to create an ensemble that contains all metastable states -- and only those -- with equal probability and to enable transitions between them, such that in a second step standard Monte Carlo techniques can be employed in order to investigate their properties.

As a first step, we set up a composite state that contains an unspecified spin configuration $\mathbf S$ and $jN$ random numbers $\{\xi\}\in(0,1]^{jN}$. Here $j$ specifies how many random numbers per lattice site are used. If basic Monte Carlo steps like spin flips and randomizations of elements of $\{\xi\}$ are applied to this state in an unbiased fashion, the system will perform simple sampling of the state space of spin configurations and simultaneously of $(0,1]^{jN}\subset\mathbb{R}^{jN}$.

Now we interpret the composite state as a random quench, i.e., the random numbers $\{\xi\}$ are used to create a sequence of spin configurations that starts at $\mathbf S$ and is guaranteed to end in a metastable state $\rho$. Applying the same Monte Carlo steps as before, changes in $\mathbf S$ and $\{\xi\}$ will, therefore, often cause changes of $\rho$ such that a random walk in the space of local minima is performed. However, it can not be expected that all metastable states are visited with equal probability.

We bias the ensemble such that a composite state is represented with a probability proportional to a weight $P_{\rm goal}({\mathbf S},\{\xi\})$. The function $P_{\rm goal}$ is chosen such that the resulting random walk in the space of local minima performs simple sampling, i.e., all metastable states $\rho$ are occupied with uniform probability.

\paragraph{The principal ensemble}

%The second algorithm is a Monte Carlo method that does not use basic spin configurations of the spin glass as its states, but instead short sequences of a primitive Monte Carlo chain. These can be specified by an initial spin configuration $\mathbf S$ and a set of random numbers $\{\xi\}$ which the primitive MC method $\mathcal{M}$ maps onto a sequence of basic states:

The initial spin configuration $\mathbf S$ and the set of random numbers $\{\xi\}$ is mapped onto a sequence of spin configurations by the primitive Monte Carlo method $\mathcal{M}$:

\begin{equation}
\mathcal{M}:(\mathbf{S},\{\xi\}) \mapsto ( \sigma_0,\sigma_1,\sigma_2,\dots,\sigma_f ),
\end{equation}
where $\sigma_i$ are spin configurations of the spin glass with $\sigma_0 \equiv \mathbf{S}$. In our case $\mathcal{M}$ stands for an energy minimization. The random numbers $\{\xi\}$ are used to randomly pick a spin with positive energy in $\sigma_i$, which is then flipped, thus creating $\sigma_{i+1}$. This is repeated until all spins have negative energy, such that the final state $\sigma_f\equiv\rho$ is a local energy minimum.

%During the overarching simulation, $\mathbf{S}$ and $\{\xi\}$ are both altered and consequently $\sigma_f$ changes in the process. We will now show how an ensemble can be defined such that all local minima are thereby visited with a uniform probability.

As usual, the primitive methods that are used to modify $\mathbf{S}$ and $\{\xi\}$ are unbiased, i.e., if simple sampling (ss) were used all $\mathbf{S}$ would be visited with equal probability and the random numbers $\{\xi\}$ would be uniformly distributed. In such a process the probability to obtain a given sequence $( \sigma_0,\dots,\sigma_f )$ can easily be determined. Since $\mathcal{M}$ does an unbiased selection from all spins with positive energy, the probability of each individual draw equals the inverse numbers of spins with positive energy and the total probability is proportional to their product:
\begin{equation}
P_{\rm ss}\left(( \sigma_0,\dots,\sigma_f )\right) \equiv P_{\rm ss}\left( \mathcal{M}(\mathbf{S},\{\xi\}) \right) = P(\mathbf{S}) \prod\limits_{i=0}^{f-1}\frac1{n_i^{\rm p}},
\end{equation}
where $n_i^{\rm p}$ is the number of spins with positive energy in the configuration $\sigma_i$. Using the inverse of this probability as the statistical weight of a state in a biased ensemble, 
\begin{equation}
P_{\rm is}( \mathbf{S},\{\xi\} ) \propto P_{\rm ss}\left( \mathcal{M}(\mathbf{S},\{\xi\}) \right)^{-1},
\end{equation}
during an importance-sampling (is) simulation will cause this simulation to create all sequences $( \sigma_0,\dots,\sigma_f )$ with equal probability.

Of course, our goal is not to sample all sequences with equal probability, but all final states $\sigma_f$, which are the local energy minima. We have to assign an additional weight $W$ to each sequence such that for all local minima $\rho$
\begin{equation}
p(\rho)=\sum\limits_{\substack{( \sigma_0,\dots,\sigma_f ) ,\\ \sigma_f=\rho }}W\left(( \sigma_0,\dots,\sigma_f )\right)={\rm const}.
\end{equation}
This is in a sense the inverse process to the first reweighting, where we introduced a weight function in order to move from a uniform distribution over the starting configurations $\mathbf{S}$ or $\sigma_0$ to a uniform distribution over all sequences. Now we wish to abandon the latter in favor for an ensemble with a uniform distribution over the final states $\sigma_f$. 

%%%%%%%%%%%%%%%%%%%%%%%%%%%%%%%%%%%%%%%%% 
\begin{figure}
\begin{center}
\includegraphics[width=.9\columnwidth]{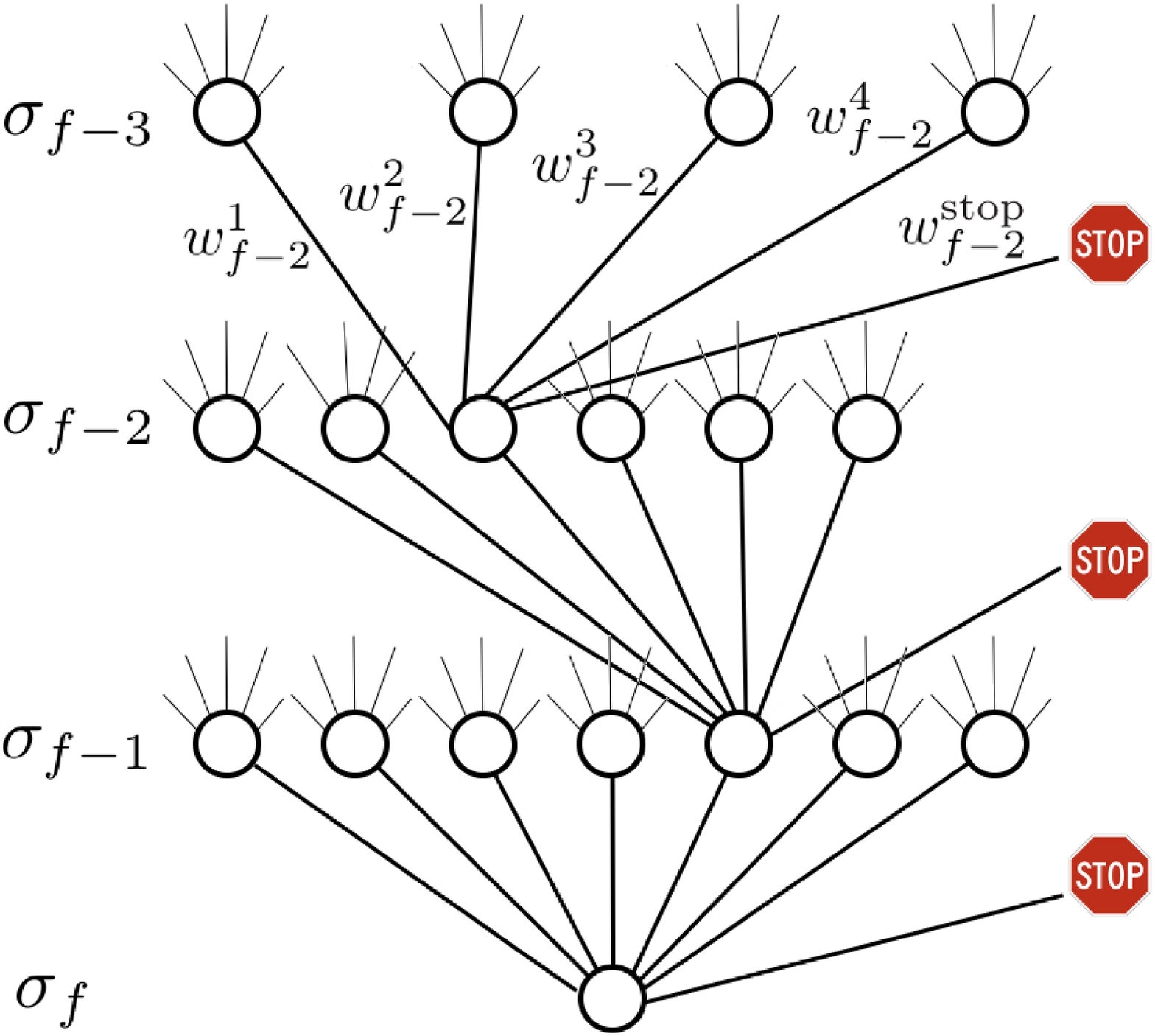}
\caption{\small{\label{fig:up_tree} \emph{Branches of the tree of sequences for one particular final state $\sigma_f$. Circles represent spin configurations and STOP-nodes indicate that the initial configuration $\sigma_0$ has been identified with the node below. See text for a detailed description.}}}
\end{center}
\end{figure}
%%%%%%%%%%%%%%%%%%%%%%%%%%%%%%%%%%%%%%%%%

Consider the partial tree depicted in Fig.~\ref{fig:up_tree}. The full tree contains all those sequences that end in one particular final state, the local minimum $\sigma_f$. Each circle represents a spin configuration and any two connected states differ by exactly one spin value with the energy decreasing towards the root. The same configuration can appear multiple times in the tree. One possibility is that spins can be flipped in a different order which will cause states to appear more than once at the same level. We reconstruct the sequences in reverse order, i.e., starting with the final state $\sigma_f$ and proceeding upwards to previous states. The length of the sequences is variable. Therefore, if all possible paths from the root $\sigma_f$ should be considered, we have to accommodate for the possibility  of a premature stop while a continuation towards configurations of higher energy is still possible. This is symbolized by the STOP-nodes. They are not configurations themselves, but identify their parent node with $\sigma_0$. If we assign weights $w$ to all branches such that the sum over the weights of a node's outgoing (upward) branches equals unity
\begin{equation}
w_i^{\rm stop}+\sum\limits_{j=1}^{n_i^{\rm n}}w_i^j=1,
\end{equation}
then the product of these weights along each path has the desired property of $W$. Here, $n_i^{\rm n}=N-n_i^{\rm p}$ is the number of spins with negative energy in the configuration $\sigma_i$ and, therefore, the number of spin configurations with higher energy to which $\sigma_i$ can be connected. The most intuitive solution is to assign equal weights to all true continuations:
\begin{equation}
w_i^j=\frac{1-w_i^{\rm stop}}{n_i^{\rm n}},\qquad j=1,2,\dots,n_i^{\rm n}.
\label{eq:tree_weight_1}
\end{equation}
The remaining weights $w^{\rm stop}$ determine the length $f+1$ of the sequence.  We chose
\begin{equation}
w_i^{\rm stop}= 
\begin{cases}
  0 & \text{if $f+1-i<l_{\min}$} \\
  \frac1{l_{\max}-f+i} & \text{else}
\end{cases}
\label{eq:tree_weight_2}
\end{equation}
in order to obtain sequences of any length between (and including) $l_{\rm min}$ and $l_{\rm max}$ with equal frequency. If, for instance, $l_{\rm min}=3$ and $l_{\rm max}=7$ the STOP-weights $w_i^{\rm stop}$ for the levels in Fig.~\ref{fig:up_tree} from the bottom to the top would read $0,0,\frac15,\frac14,\frac13,\frac12,1$.
We find 
\begin{equation}
W\left(( \sigma_0,\dots,\sigma_f )\right)=w_0^{\rm stop} \prod\limits_{i=1}^{f}\frac{1-w_i^{\rm stop}}{n_i^{\rm n}},
\end{equation}
which with our choice of $w_i^{\rm stop}$ simplifies to
\begin{eqnarray}
W\left(( \sigma_0,\dots,\sigma_f )\right) &=& \frac1{l_{\rm max}-l_{\rm min}+1} \prod\limits_{i=1}^{f}\frac1{n_i^{\rm n}},\\
 &\propto& \prod\limits_{i=1}^{f}\frac1{n_i^{\rm n}}
\end{eqnarray}
if $f+1\in\{l_{\rm min},\dots,l_{\rm max}\}$, otherwise $W\left(( \sigma_0,\dots,\sigma_f )\right)=0$.

Eventually, multiplying $P_{\rm is}( \mathbf{S},\{\xi\} )$ and $W( \mathbf{S},\{\xi\} )$, we are left with an ensemble which includes all final states $\sigma_f$ with equal probability
\begin{equation}
P_{\rm goal}( \mathbf{S},\{\xi\} ) \coloneqq \frac{ \prod\limits_{i=0}^{f-1}n_i^{\rm p} }
                                                 { \prod\limits_{i=1}^{f}n_i^{\rm n} }.
\end{equation}
The freedom to restrict the length of the sequence from above is important since during the construction of $W$ we implicitly assumed that we can always move to states with higher energy and could, therefore, always chose $w^{\rm stop}\ne 1$. To ensure that this assumption is justified during a simulation the sequences must not be too long. In our simulations we chose $l_{\rm max}=N/3$ and $l_{\rm min}=1$.

\paragraph{The minimization}

As stated above the minimization method $\mathcal{M}$ randomly chooses spins with positive energy and flips these until a stable configuration is reached. Normally, in such a procedure a spin would be selected by considering all candidates and using a single random number uniformly distributed between zero and the number of spins with positive energy, which rounded up will determine which spin to flip. However, this kind of global selection is unsuited in our case. It would cause a modification of the initial state $\sigma_0$ to potentially affect every single selection which would, therefore, require a complete reconstruction of the $\sigma_i$. The changes to $\sigma_f$ would be considerable, which is undesirable in a Monte Carlo simulation, since the resulting acceptance rates would be very small. Instead, we implement strictly local conditions which collectively effect a global selection.

In the following as we construct the sequence $\sigma_0,\sigma_1,\dots,\sigma_f$ we will record the evolution of single spins by means of `spin states' $\zeta$ which describe a spin's value and its energy as determined by the environment, i.e., the adjacent spins. The initial state $\zeta_{0,k}$ of spin $s_k$ as given by the spin configuration $\sigma_0$ will change to a new state $\zeta_{1,k}$ as soon as its energy changes, which happens if either spin $s_k$ or one of its neighboring spins get flipped. Since consecutive $\sigma_i$ only differ in exactly one spin, the first index of the $\zeta$ will in general not agree with the spin configurations to which they belong. For instance, most states $\zeta_{0,k}$ will be shared by $\sigma_0$ and $\sigma_1$.

If we assign a uniformly distributed random number $\eta_{0,i}\in(0,1]$ to each initial spin state $\zeta_{0,i}$ with positive energy and sort the spins based on the magnitude of $\eta_{0,i}$ it is clear the we will obtain a completely random sequence. We identify the largest $\eta_0$ and flip the associated spin, thus creating $\sigma_1$. We then proceed with the new largest $\eta$. The following rules apply:
\begin{itemize}
\item{If by the flip of a spin with random number $\eta_{\mu,k}$ an adjacent previously stable spin acquires positive energy in its new state $\zeta_{\nu,l}$, it can easily be inserted into the ordered set of spins with positive energy by choosing the random number $\eta_{\nu,l}\in(0,\eta_{\mu,k}]$.}
\item{Albeit not strictly necessary, a new random number is also assigned in the same way if the energy of an unstable spin changes but remains positive.}
\item{If a spin with positive energy changes to a stable state, i.e., with negative energy, its $\eta$ is removed from the ordered set.}
\item{The non-adjacent spins that retain their energy during a spin flip and whose state therefore does not change keep their random number.}
\end{itemize}
For reasons of efficiency, in our simulation we reserve one random number $\xi\in(0,1]$ for each spin state regardless of the sign of its energy and calculate $\eta$ from it when required. Random numbers of spins with negative energy have no impact and can be considered as temporarily decoupled degrees of freedom.

We can reformulate the algorithm by introducing a flipping `time'
\begin{equation}
t_{\mu,k}\coloneqq-\ln\eta_{\mu,k}
\label{eq:def_time}
\end{equation}
and consequently
\begin{equation}
t_{\mu,k}=t^{\rm orig}_{\mu,k}-\ln\xi_{\mu,k},
\label{eq:def_time2}
\end{equation}
where $\xi_{\mu,k}$ is a uniformly distributed random number $\in(0,1]$ and $t^{\rm orig}_{\mu,k}$ is the time at which the particular state of spin $s_k$ was created, i.e., the time when a flip of one of its neighbors last changed its energy. A flip of a spin will not lead to a new flipping time for itself since after a flip its energy is per definition negative. In the beginning, i.e., for the state $\sigma_0$ all $t^{\rm orig}_{0,k}=0$. Although not used in our work, it is instructive to consider a biased selection. If the spin $s_k$ in state $\zeta_{\mu,k}$ shall be selected with the relative weight $v_{\mu,k}$ it is relatively easy to show that
\begin{equation}
t_{\mu,k}=t^{\rm orig}_{\mu,k}-\frac{\ln\xi_{\mu,k}}{v_{\mu,k}}
\end{equation}
generates the desired behavior. If weights proportional to the Boltzmann weight of the respective energy change were chosen and if energy increases were allowed $\mathcal{M}$ would become the waiting-time Monte Carlo method.\cite{WTM}

We will now briefly discuss the different methods that we use in order to modify the sequence $\{\sigma\}$ during out Monte Carlo simulation.

\paragraph{Top-down update}

If instead of the spin with the largest random number $\eta$ (or the smallest flipping time $t$) the spin with the highest energy were flipped, the thus derived deterministic method would constitute a so-called greedy algorithm. Both methods are structurally very similar, each flip only affects adjacent spins and each spin state is characterized by a quantity (energy or $\eta$) whose maximum determines the next step. There is one difference that makes the random minimization easier to handle. During the greedy algorithm within the sequence of flipped spins the energy sometimes increases, while in the method used here $\eta$ will always decrease ($t$ will always increase). In a recent article \cite{hoga} about `dynamical greedy algorithms' for the Edwards-Anderson model we have discussed algorithms that will propagate changes in the initial configuration $\sigma_0$ and determine the new $\sigma_f$ with little computational effort. Since the same ideas are used here with very little modification to implement a method that allows for changes to $\mathbf S$, i.e., to $\sigma_0$ we refer to this publication for details.

Since random changes to the initial configuration will in general change the weight $W_{\rm goal}$ the acceptance probability must contain the ratio
\begin{equation}
R = \frac{ P_{\rm goal}( \mathbf{S}_{\rm new},\{\xi_{\rm new}\} ) }{ P_{\rm goal}( \mathbf{S}_{\rm old},\{\xi_{\rm old}\} ) }
\end{equation}
as a bias correction.

\paragraph{Single random number update}

Besides the starting configuration we can and should also modify the random numbers $\xi$ during the simulation. All random numbers that belong to spin states with negative energy can be updated at leisure and those which belong to spin states $\zeta_{\mu,k}$ with positive energy, but which do not lead to a spin flip since the state is replaced before the flipping time:
\begin{equation}
t_{\mu,k}>t^{\rm orig}_{\mu+1,k}
\end{equation}
can be assigned a new uniformly distributed random number
\begin{equation}
\xi_{\mu,k}\in\left[0, \exp({-t^{\rm orig}_{\mu+1,k}})\right).\
\label{eq:upd_xi}
\end{equation}
In both cases the sequence of spin flips and therefore the weight $W_{\rm goal}$ remains unchanged. Hence, these updates can always be accepted with probability one. It is, of course, possible to modify any $\xi$ without constraint and proceed to determine the resulting possibly altered sequence $\sigma_0,\dots,\sigma_j,\sigma_{j+1}',\dots,\sigma_f'$. Then a non-trivial acceptance probability would ensue. In our simulations, however, we do not use such a method.

\paragraph{Bottom-up update}
The distribution of sequences as defined by (\ref{eq:tree_weight_1}) and (\ref{eq:tree_weight_2}) enables us to introduce another update. Exploiting the fact that all sequence lengths between $l_{\rm min}$ and $l_{\rm max}$ are equally likely just as any upward continuation in Fig.~\ref{fig:up_tree}, it is possible to create a new sequence that ends with the old local minima $\sigma_f$:
\begin{itemize}
\item{ Chose a new length $f'+1$ from the allowed values randomly. }
\item{ Starting from $\sigma'_{f'}=\sigma_f$ create $\sigma'_{f'-1},\sigma'_{f'-2},\dots,\sigma'_0$ by randomly flipping spins with negative energy. }
\item{ Assign new random numbers $\{\xi\}$ that are consistent with this sequence. }
\end{itemize}
The third point warrants a more thorough discussion. Naturally, we start with $\sigma'_0$ since the random numbers $\eta'_0$ do not depend on the latter ones. If $\sigma'_0$ possesses $n^{\rm p}_0$ spins with positive energy and if $\sigma'_1$ is reached by flipping $s_k$ we have to assign $\eta_{0,k}$ according to a distribution that equals the distribution of $\max(\chi_1,\chi_2,\dots,\chi_{n^{\rm p}_0})$, where the random numbers $\chi_i\in(0,1]$ are uniformly distributed. We find the distribution
\begin{equation}
p\left( \eta_{0,k} \right)= n^{\rm p}_0 (\eta_{0,k})^{n^{\rm p}_0-1},\quad \eta_{0,k}\in(0,1]
\end{equation}
which means that we can set
\begin{equation}
\eta_{0,k}=\chi^\frac1{n^{\rm p}_0},
\end{equation}
where $\chi\in(0,1]$  is uniformly distributed. Similarly, if the spin flip from $\sigma'_{i-1}$ to $\sigma'_i$ occurred at time $\tau_i$ and if the spin $s_l$ has to be flipped in order to reach the new state $\zeta_{\mu+1,l}$ and $\sigma'_{i+1}$, $\eta_{\mu,l}$ is distributed the same way as $\max(\chi'_1,\chi'_2,\dots,\chi'_{n^{\rm p}_i})$, with $\chi'_i\in(0,e^{-\tau_i})$ uniformly distributed. Hence
\begin{equation}
p\left( \eta_{\mu,l} \right)= n^{\rm p}_i \left( \eta_{\mu,l} \right)^{n^{\rm p}_i-1}(e^{\tau_i})^{n^{\rm p}},\quad \eta_{\mu,l}\in(0,e^{-\tau_i})
\end{equation}
and therefore we can calculate $\eta_{\mu,l}$ from a uniformly distributed random number:
\begin{equation}
\eta_{\mu,l}=\chi^\frac1{n^{\rm p}_i}e^{-\tau_i}.
\end{equation}
Once all times and respective random numbers $\eta$ of the performed flips are defined, their basic random numbers $\xi$ can be calculated for the known times by inverting (\ref{eq:def_time2}) and for the remaining spin states with positive energy according to (\ref{eq:upd_xi}). All other $\xi$ are decoupled and may be kept or chosen at random. Since the update is designed to create sequences with the desired distribution there is no bias to correct by the acceptance probability:
\begin{equation}
R = 1.
\end{equation}

It is worth noting that in principle this update allows for a true Markovian chain in the space of local minima. Without it, the selection of a new state $\rho'$ does not exclusively depend on the current state $\rho$, but on the hidden degrees of freedom in $\mathbf S$ and $\{\xi\}$. A Markovian process is performed in their state space. Now, we can completely randomize these hidden degrees of freedom after each step using the bottom-up update, thus removing the surplus `memory' from the system. However, since the procedure involves the entire system and is, therefore, computationally expensive, it is not advisable to apply it that often.

In our simulations we randomly select a spin in $\mathbf S$ and attempt a spin flip in a top-down update. Then, the random variables at this lattice site are updated if possible. After $N$ such combinations we perform a single bottom-up update.

\paragraph{Simulation}

Once the framework introduced above is in place we can ignore all its inner workings and treat it as a normal system which ergodically changes from one single-flip stable configuration $\rho\equiv\sigma_f$ to another. Of course, the configurations of this particular system are a subset of the states of another system, but this concerns us no longer. In order to obtain statistics for a large energy range we apply a flat-histogram method. We introduce another weight function 
\begin{equation}
W_{\rm flat}(E) \approx \Omega(E)^{-1}
\end{equation}
and require that in our simulation the probability to visit a certain metastable state $\rho$ is
\begin{equation}
P_{\rm flat}(\rho) \propto W_{\rm flat}\left( E(\rho)\right),
\end{equation}
which means that new states are accepted with the probability
\begin{equation}
P_{\rm flat}^{\rm acc}(\rho_{\rm old}\rightarrow \rho_{\rm new}) = \min\left( 1 , \frac{ W_{\rm flat}\left( E(\rho_{\rm new})\right) }{ W_{\rm flat}\left( E(\rho_{\rm old})\right) } R \right).
\end{equation}
We initially approximate $W_{\rm flat}$ using a variant of the well-known Wang-Landau algorithm \cite{Wang_Landau} with an additional restriction. The algorithm has difficulties to converge and to sample the distribution in the extreme tails, i.e., at low and at high energy because only very few states exist there. The high-energy minima are much harder to find than the low-energy ones. We suspect the reason is that the latter are embedded in large basins and large metabasins which help to guide the simulation. The problematic regions can be excluded from the simulation by restricting $W_{\rm flat}$:
\begin{equation}
W_{\rm flat}(E)< 
\begin{cases}
  \min(W_{\rm flat}(E))+\Delta_{\rm L} & \text{if $E<E^*$} \\
  \min(W_{\rm flat}(E))+\Delta_{\rm R} & \text{if $E>E^*$},
\end{cases}
\end{equation}
where $E^*$ is the position of the minimum of $W_{\rm flat}$
\begin{equation}
W_{\rm flat}(E^*)  = \min(W_{\rm flat}(E)).
\end{equation}
This is more convenient than restricting the energy range directly because it can be applied in the same way to all samples. The values that we use are listed in Table~\ref{tab:w_restr}. Choosing $\Delta_{\rm L}=0.205N$ still allows for the sampling of the ground state, but prevents the algorithm to spend too much time at low energies during the weight determination.

Once $W_{\rm flat}(E)$ is known with sufficiently high precision, we perform the main simulation and record a histogram $H(E)$ of the local minima from which their distribution can be calculated:
\begin{equation}
\Omega(E)=\frac{H(E)}{W_{\rm flat}(E)} \, .
\end{equation}

\begin{table}
\caption{\small{\label{tab:w_restr} \emph{Upper bounds for the weight function $W_{\rm flat}(E)$.}}}
\begin{center}
\begin{tabular}{c c c}
\toprule
$L$ & $\Delta_{\rm L}$ & $\Delta_{\rm R}$\\  \hline
 4  &  $\infty$ &  $\infty$\\
 6  & $\infty$ & $0.18L^3$\\
 8  & $0.205L^3$ & $0.17L^3$\\
 10  & $0.205L^3$ & $0.13L^3$\\
\botrule
\end{tabular}
\end{center}
\end{table}

%%%%%%%%%%%%%%%%%%%%%%%%%%%%%%%%%%%%%%%%%%%%%%%%%%%%%%%%%%%%%%%%%%%%%%%%%%%%%%%%%%%%%%%%%%%%%%%%%%%%%%%%%%%%%%%%%%%%%%%%%%%%%%%%%%%%%%%%%%%%%%%%%%%%%%%%%%%%%%%%%%%%%%%%%%%%%%%%%%%%%%%%%%%%%%%%%%%%%%%%%%

\section{Results}

As a first goal we test the validity of our methods. 
%%%%%%%%%%%%%%%%%%%%%%%%%%%%%%%%%%%%%%%%% 
\begin{figure}
\begin{center}
\includegraphics[width=.9\columnwidth]{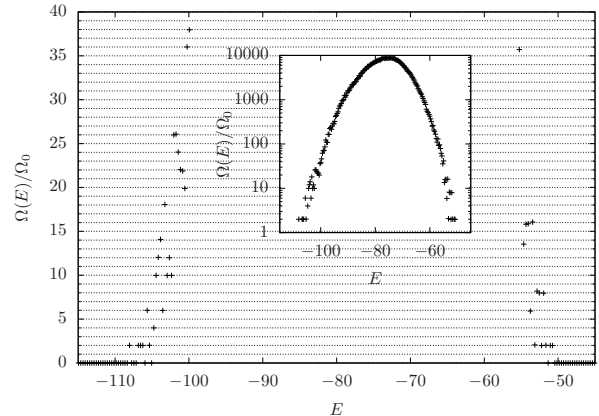}
\caption{\small{\label{fig:dos_int} \emph{
%The suitably normalized distribution of local minima for a single $4\times4\times4$ sample of the Edwards-Anderson model as measured with method II. Data cluster around integer multiples.
The tails of the distribution of local minima for a single $4\times4\times4$ sample of the Edwards-Anderson model as measured with method II. The complete distribution is shown in the inset. The $\Omega$-values are integer multiples of the lowest non-zero value. The appropriate normalization $\Omega_0$ is not obtained from the simulation, but is deliberately chosen such that the lowest occupied level equals $2$ indicating that the respective energy intervals each contain a single twofold degenerate metastable state.
}}}
\end{center}
\end{figure}
%%%%%%%%%%%%%%%%%%%%%%%%%%%%%%%%%%%%%%%%%
In Fig.~\ref{fig:dos_int} we show $\Omega(E)$ for a $L=4$ sample of the Edwards-Anderson model measured with method II. We use a binning method, i.e., every data point represents the aggregated statistics from a small energy interval. Each interval contains an integer number of local minima, hence a measurement of $\Omega$ should produce values that are integer multiples of the lowest non-zero value. This becomes clearly apparent in the tails of the distribution. We interpret the larger statistical fluctuations at high energy as evidence that sampling the high-energy minima is more demanding.
%%%%%%%%%%%%%%%%%%%%%%%%%%%%%%%%%%%%%%%%% 
\begin{figure}
\begin{center}
\includegraphics[width=.9\columnwidth]{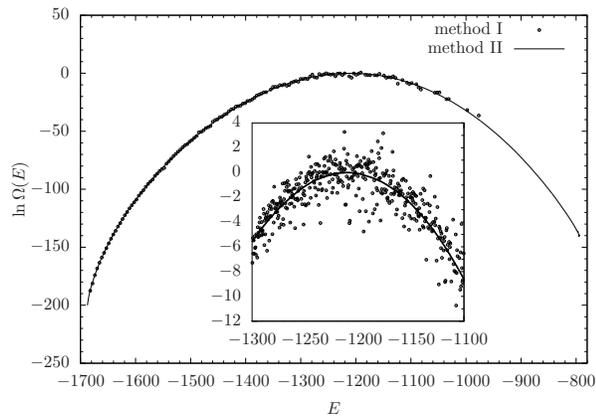}
\caption{\small{\label{fig:method_comp} \emph{The distribution of local minima for a single $10\times10\times10$ sample of the Edwards-Anderson model measured with both methods. In the main plot only every tenth data point for method I is displayed in order to ensure visibility. Notwithstanding, both data sets have a similar resolution. While the results are consistent, which suggests that both methods are accurate, the statistical error of method II is substantially lower (see inset). Besides, at high energy method I failed to find minima in a number of energy intervals leading to an apparent thinning-out of data points.}}}
\end{center}
\end{figure}
%%%%%%%%%%%%%%%%%%%%%%%%%%%%%%%%%%%%%%%%%

In order to compare both methods we show results for a $10\times10\times10$ system in Fig.~\ref{fig:method_comp}. We find that the results are in agreement. However, method I suffers from a much larger statistical error and for larger systems method II is able to cover a much larger energy range. Consequently, we proceed to apply method II if possible, i.e., for the Edwards-Anderson model.

\subsection{Total number of minima}

We note that the raw data from the simulations are a priori not normalized. Only for $L=4$ and method II it is in principal possible to obtain a normalization constant since very precise measurements in the tails of the distribution are  required. This was done `manually' for the distribution in Fig.~\ref{fig:dos_int}. To automatize the normalization for small systems we use the lower values of the distribution of minima $\Omega$ to define a fitness function
\begin{equation}
\mathcal{F}(s)=\sum\limits_{\substack{i ,\\ 0<\Omega(E_i)<S}}\frac{\cos\left( 2\pi\Omega(E_i)/s \right)}{\Omega(E_i)}\quad,
\end{equation}
which is intended to quantify how well the data matches a supposed level distance $s$. Here, the parameter $S$ is a suitably chosen upper threshold that limits the computational effort and increases precision while the division by $\Omega(E)$ is introduced since we presume that $\ln\Omega(E)$ has a constant statistical error which implies that $\Omega(E)$ has a statistical error proportional to its absolute value and that larger values should, therefore, contribute less. The true level distance, i.e., the statistical contribution of a single (twofold degenerate) local minimum is then estimated by the position of the maximum of $\mathcal{F}(s)$ and dividing by 2 accounts for the degeneracy:
\begin{equation}
\Omega_0=\argmax_{s>s_{\rm min}}\mathcal{F}(s)/2,
\label{eq:single_min_contr}
\end{equation}
where we considered
\begin{equation}
  s_{\rm min}=\frac{\min(\Omega(E_i))}{10}
\end{equation}
sufficient. It is highly unlikely that for $L=4$ and the interval width we used ($\Delta E=E_{i+1}-E_i=0.3J$) the interval with the lowest population already contains more than 10 local minima. We can now properly normalize the distributions of local minima and determine the total number of minima
\begin{equation}
%N_S=\int_{-\infty}^\infty\Omega(E)/\Omega_0 dE.
N_S=\sum_{i}^\infty\Omega(E_i)/\Omega_0,
\end{equation}
%The result is displayed in Fig.~\ref{fig:n_min_hist}.
%%%%%%%%%%%%%%%%%%%%%%%%%%%%%%%%%%%%%%%%% 
%\begin{figure}
%\begin{center}
%\includegraphics[width=.9\columnwidth]{./n_min_hist.eps}
%\caption{\small{\label{fig:n_min_hist} \emph{The number of samples as a function of the logarithm of the number of local minima for $L=4$.}}}
%\end{center}
%\end{figure}
%%%%%%%%%%%%%%%%%%%%%%%%%%%%%%%%%%%%%%%%%
take the disorder average, and obtain an estimate for the number\footnote{Here, we have used the logarithm of the average of $N_S$ in order to be able to compare our result. However, since both $N_S$ and $\Omega(E)$ are expected to be log-normal distributed, in the rest of the paper we  consider averages of logarithmic quantities.} of metastable states for the three-dimensional EA model with $L=4$:
\begin{equation}
%\langle\ln N_S\rangle_J/N=0.2100(4)
\ln\langle N_S\rangle_J/N=0.2111(3).
\end{equation}
This result matches the value $0.21125(1)$ that was calculated in Ref.~\onlinecite{Waclaw_Burda} using an analytic approximation.

\subsection{Averaging}

Unfortunately, for larger system this normalization method can not be applied and all other distributions presented henceforth are only determined up to unknown factors. Since we calculate the average of the logarithm of the distributions, these become unknown additive constants which do not have a direct effect besides creating an unknown additive constant for the average as well. However, the second moments $\langle \left(\ln\Omega(E)\right)^2 \rangle_J$ and hence the estimators of the statistical errors of $\langle \ln\Omega(E) \rangle_J$ depend on these constants. In our analysis we chose them such that the maximum of the canonical distributions $\Omega(E)e^{-\beta_{\rm sync}E}$ is identical for all samples. This is equivalent to the (not entirely valid) assumption that all samples have about the same number of local minima and leads to an underestimation of the statistical error. With increasing system size this effect will vanish. Since we obtain very precise data for the Edwards-Anderson model we can use the natural choice $\beta_{\rm sync}=0$. For the SK model, however, the data is very noisy around the maximum of $\Omega(E)$ and we use $\beta_{\rm sync}=0.05$ for $N=96$ and $\beta_{\rm sync}=0.2$ for $N=128$. The first averaging procedure is given by
\begin{equation}
[\ln\Omega(E)]_1 \coloneqq \langle\, \ln\Omega(E)-\ln\Omega(E^*)\,\rangle_J,
\end{equation}
where $E^*$ is the position of the maximum of $\Omega(E)e^{-\beta_{\rm sync}E}$.

Due to the variability of the interactions $J_{ij}$ the energy interval at which local minima exist shifts especially for small systems. This means that  we can obtain data from all samples only from a relatively small energy region. Outside this interval no average can be computed since the logarithm of the missing distributions is not defined. To obtain an averaged function over a larger interval we introduce a second averaging procedure for which we shift all distributions along the energy axis such that their maxima coincide with the average maximum position:
\begin{equation}
[\ln\Omega(E)]_2 \coloneqq \left\langle\, \ln\Omega(E+E^*-\langle E^*\rangle_J)-\ln\Omega(E^*)\,\right\rangle_J.
\end{equation}

\subsection{Sherrington-Kirkpatrick model}
We simulate systems of size $N=48,64,96,128$ with method I using parameters according to Table~\ref{tab:sk_param}. For each size we investigated 200 samples.
\begin{table}
\caption{\small{\label{tab:sk_param} \emph{Parameters used for the simulation of the SK model.}}}
\begin{center}
\begin{tabular}{c c c}
\toprule
$N$ & $\beta_{\rm min}$ & $\beta_{\rm max}$ \\  \hline
 $48$  & $-1.44$  & $1.4$ \\
 $64$  & $-1.4$  & $1.4$ \\
 $96$  & $-0.6$  & $1.4$ \\
 $128$  & $-0.4$  & $1.4$ \\
\botrule
\end{tabular}
\end{center}
\end{table}

\begin{figure}
\begin{center}
\includegraphics[width=.9\columnwidth]{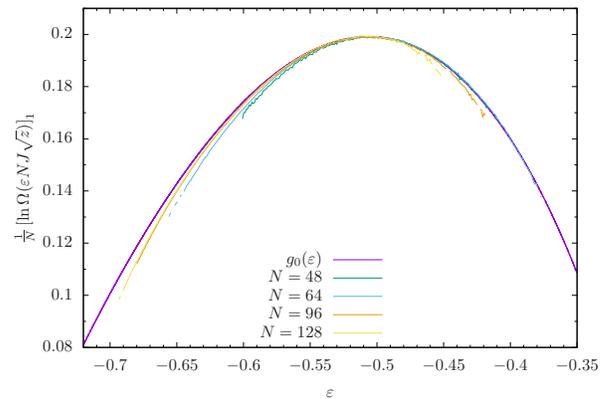}
\caption{\small{\label{fig:SK_dos_av_1} \emph{The average 1 logarithmic density of local minima for the SK model. The vertical position of the curves, i.e., the normalization of $\Omega$ have been chosen such that the maxima of the curve coincide.}}}
\end{center}
\end{figure}

\begin{figure}
\begin{center}
\includegraphics[width=.9\columnwidth]{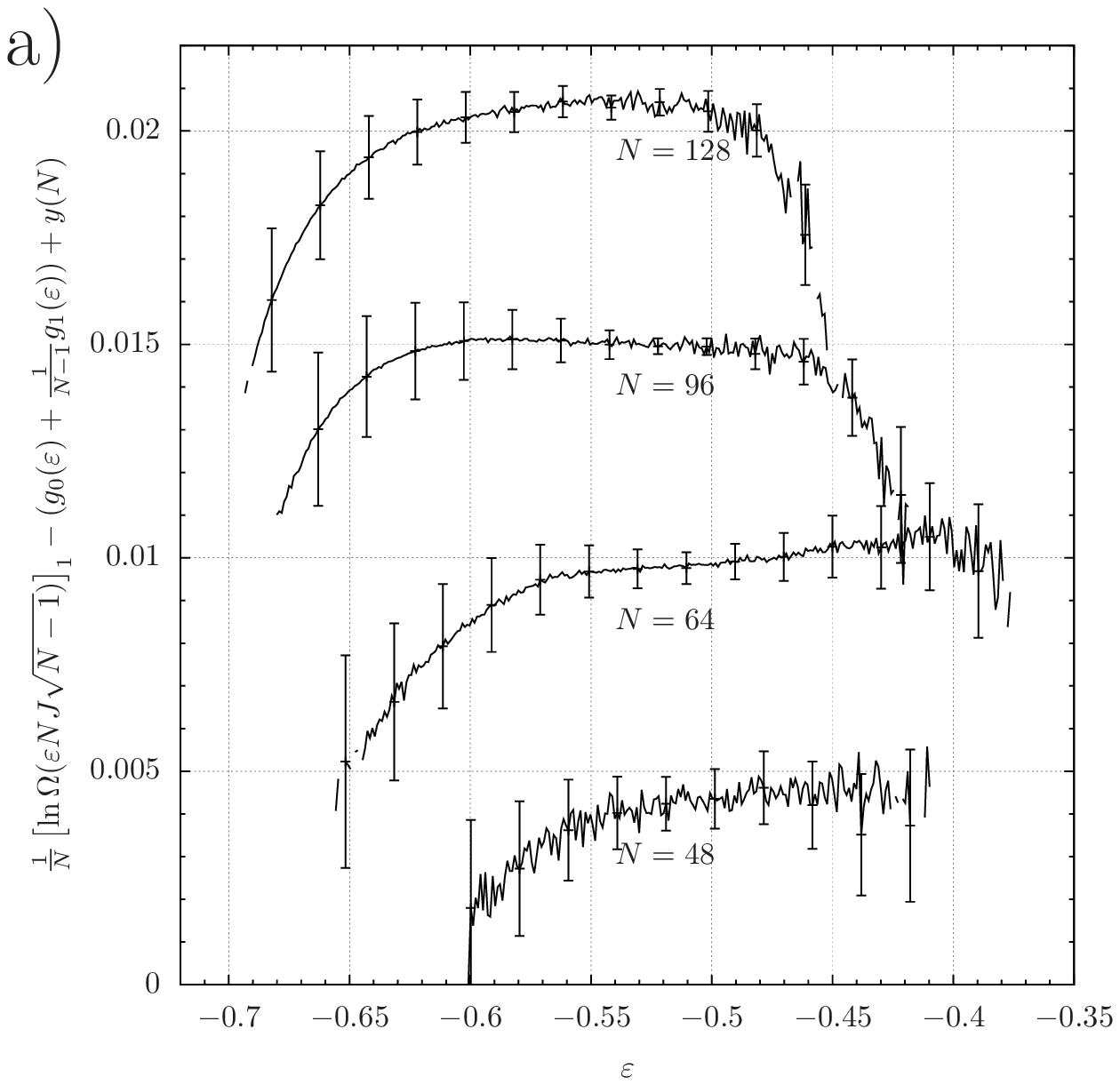}
\includegraphics[width=.9\columnwidth]{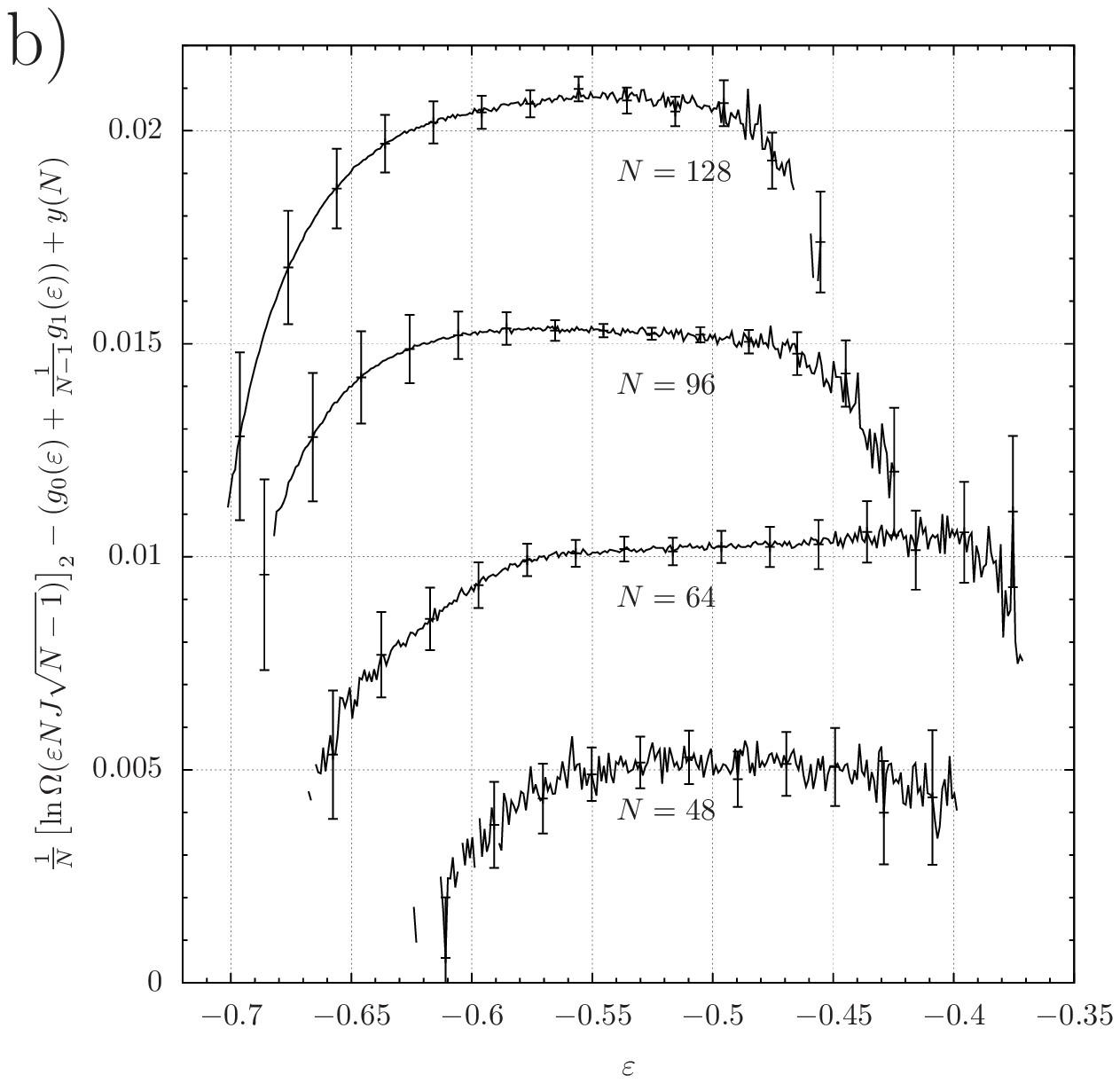}
\caption{\small{\label{fig:SK_dos_av_1_diff_g} \emph{Deviation of the averaged distribution from the analytical prediction for the SK model using a) averaging method 1 and b) method 2. The vertical positions $y(N)$ correspond to the unknown normalization constants and have here been chosen to avoid overlapping curves.}}}
\end{center}
\end{figure}

Figure~\ref{fig:SK_dos_av_1} shows the average logarithmic distribution of local minima for the first averaging method and indicates that our results are basically in agreement with the analytical prediction. Details are discernible in Fig.~\ref{fig:SK_dos_av_1_diff_g} where we show the deviation of the averages from the finite-system approximation $g_0(\varepsilon)+\frac1zg_1(\varepsilon)$ with $z=N-1$. Since we do not have a valid normalization and can not determine the correct vertical position of the curves in Fig.~\ref{fig:SK_dos_av_1} the absolute differences $\frac1N \left[ \ln\Omega(\varepsilon NJ\sqrt{N-1}) \right]-(g_0(\varepsilon)+\frac1{N-1} g_1(\varepsilon))$ have no meaning. Only the relation between the differences is relevant, i.e., a horizontal curve means agreement with the analytical prediction while a large slope indicates deviation. Consequently, the vertical positions of the curves have been adjusted for convenience and are not the result of a physically motivated normalization. Error bars result from the disorder average and represent two standard deviations.

With increasing system size the range in energy $\varepsilon$ where local minima exist expands. Regardless of the averaging technique used the distribution $\Omega(\varepsilon NJ\sqrt{N-1})$ reaches smaller and greater $\varepsilon$ for $N=64$ than for $N=48$. However, for even larger system the shortcomings of the Monte Carlo method and the increasing complexity of the energy landscape make it impossible to find minima of high energy. In fact the downward curve at high $\varepsilon$ for $N=96$ and $N=128$ suggest that even for the energies where we can find minima large populations are not accessed. The alternative explanation, that the analytical prediction is not accurate, seems less likely, especially since we obtain nice horizontal curves for $N=48$ and $N=64$. At low energies the measured distributions divert from the analytical solution as predicted by Bray and Moore. The deviation is clearly visible for $\varepsilon < -0.6$. However, the errors are relatively large and it is difficult to judge whether the  calculated $\varepsilon_c \approx -0.672$ will be realized for larger systems.

%%%%%%%%%%%%%%%%%%%%%%%%%%%%%%%%%%%%%%%%% 
%%%%%%%%%%%%%%%%%%%%%%%%%%%%%%%%%%%%%%%%%

%%%%%%%%%%%%%%%%%%%%%%%%%%%%%%%%%%%%%%%%% 
%%%%%%%%%%%%%%%%%%%%%%%%%%%%%%%%%%%%%%%%%

%%%%%%%%%%%%%%%%%%%%%%%%%%%%%%%%%%%%%%%%% 
%\begin{figure}
%\begin{center}
%\includegraphics[width=.9\columnwidth]{./SK_dos_av_2_diff_g01.eps}
%\caption{\small{\label{fig:SK_dos_av_2_diff_g01} \emph{Deviation of the average 2 distribution from the analytical prediction. The vertical positions $y(N)$ have been chosen for convenience.}}}
%\end{center}
%\end{figure}
%%%%%%%%%%%%%%%%%%%%%%%%%%%%%%%%%%%%%%%%%

\subsection{Edwards-Anderson model}

\begin{figure}[t]
\begin{center}
\includegraphics[width=.95\columnwidth]{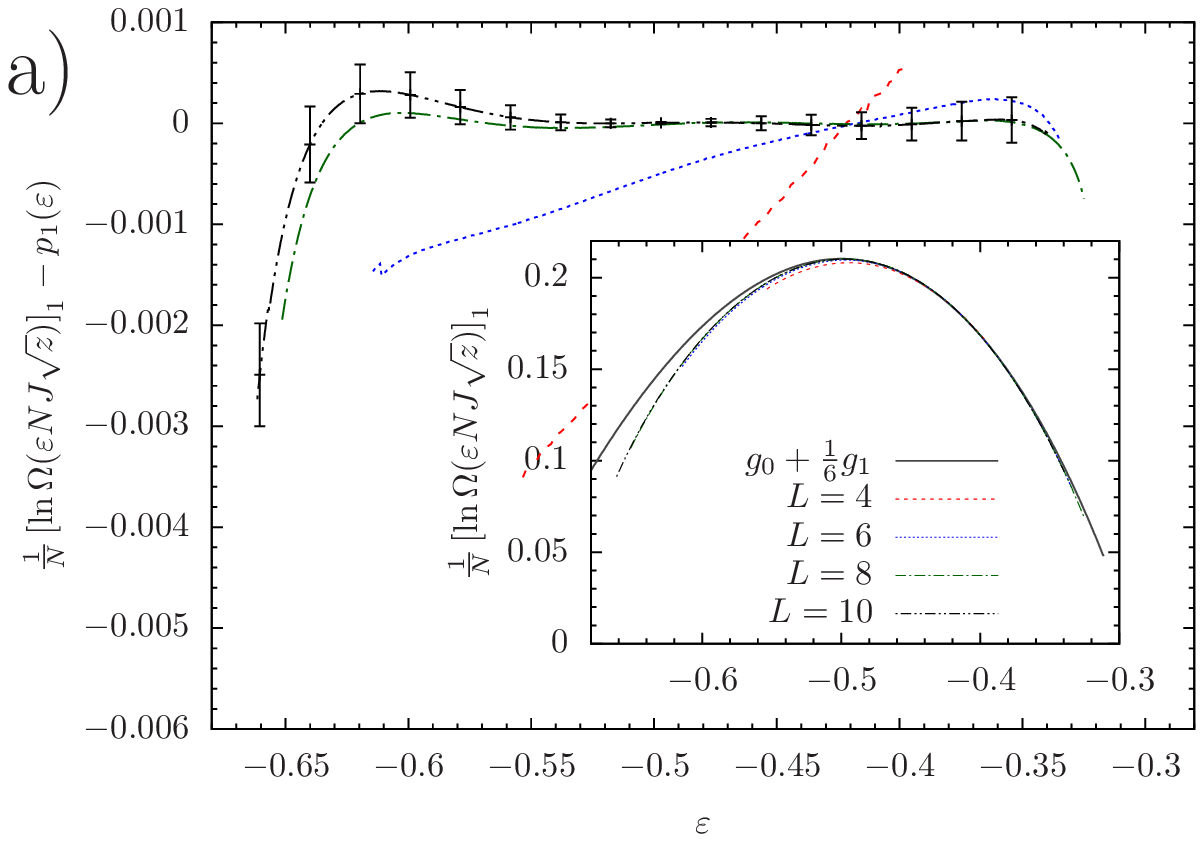}
\includegraphics[width=.95\columnwidth]{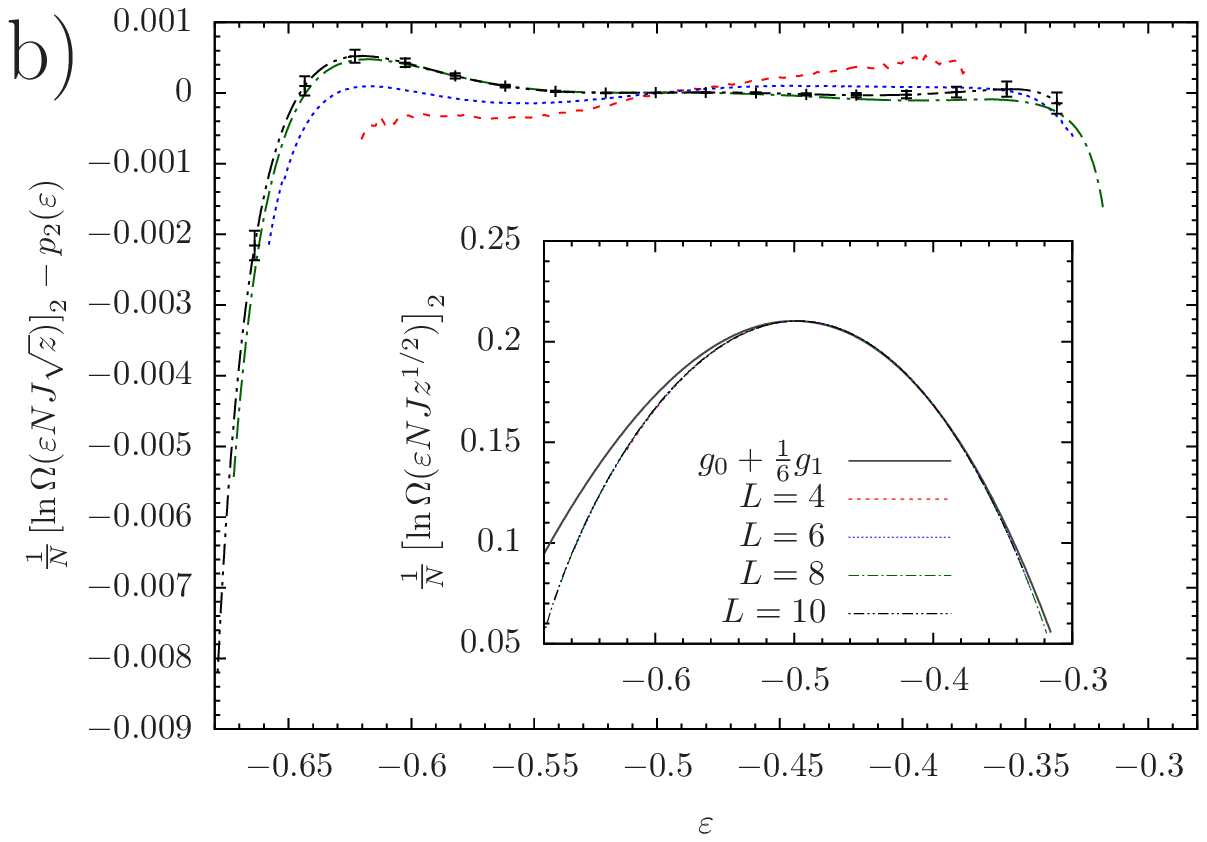}
\caption{\small{\label{fig:EA_dos_av} \emph{The logarithmic density of local minima for the EA model with the sample average taken with a) method 1 and b) method 2. The main plots show the deviations from the polynomial approximations $p_1$ and $p_2$ (see text) in order to highlight size-dependent behavior. In the insets the distributions are plotted together with the analytic approximation by Bray and Moore.}}}
\end{center}
\end{figure}

For the investigation of the Edwards-Anderson model we are able to use method II and we obtain very precise results. We consider lattices of linear extension $L=4,6,8,10$ and investigated 1000 samples for each size. In the insets of Fig.~\ref{fig:EA_dos_av} the distributions calculated with both averaging methods are plotted together with the approximation  $g_0(\varepsilon)+\frac16g_1(\varepsilon)$. We observe almost no dependence on the system size, except for the fact that the support becomes broader. Again, for large systems the sampling of local minima with high energies becomes very difficult, which led us to use the restrictions listed in Table~\ref{tab:w_restr}. Consequently, the rescaled energies reached for $L=10$ are not as high as for $L=8$. While the data agrees reasonably well with the analytical solution on the right flank of the distribution and the maximum is in a similar position, we see considerable deviations for energies below the peak. We find that the data is much better described by polynomials
\begin{equation}
p_1( \tilde{\varepsilon} )= -13.39\,\tilde{\varepsilon}^4-1.10\,\tilde{\varepsilon}^3-4.143\,\tilde{\varepsilon}^2+{\rm const}
\end{equation}
and
\begin{equation}
p_2( \tilde{\varepsilon} )= -13.73\,\tilde{\varepsilon}^4-1.07\,\tilde{\varepsilon}^3-4.142\,\tilde{\varepsilon}^2+{\rm const},
\end{equation}
where $\tilde{\varepsilon}$ is a shifted energy such that $\tilde{\varepsilon}=0$ at the maximum position for $L=10$:
\begin{equation}
\tilde{\varepsilon} = \varepsilon - (-0.4978).
\end{equation}

%\begin{equation}
%\tilde{\varepsilon} = \varepsilon - (-0.497720513193)
%\end{equation}
Both polynomials where obtained by fitting to the $L=10$ averages for $\varepsilon\in[-0.55,\infty)$. Note that the contributions from the third- and fourth-order term are small and for $\varepsilon\in[-0.6,-0.4]$ the quadratic term alone provides a very good approximation. The deviations from these polynomials depicted in the main plots in Fig.~\ref{fig:EA_dos_av} are very small and while with the first averaging method a clear dependence on the system size emerges, the curves for $L>4$ for the second averaging method are much closer together. From the statistical errors which are shown for the $L=10$ curves it also becomes clear that the second averaging method produces more precise results. We expect that both techniques would deliver the same curves for very large systems, hence we conclude that $p_2(\varepsilon)$ provides a better approximation of the true distribution of minima for large systems.

%%%%%%%%%%%%%%%%%%%%%%%%%%%%%%%%%%%%%%%%% 
%%%%%%%%%%%%%%%%%%%%%%%%%%%%%%%%%%%%%%%%%

%%%%%%%%%%%%%%%%%%%%%%%%%%%%%%%%%%%%%%%%% 
%\begin{figure}
%\begin{center}
%\caption{\small{\label{fig:EA_dos_av_2} \emph{Inset: The average 2 logarithmic density of local minima for the EA model. Main: Deviation from the polynomial approximation $p_2$ (see text). }}}
%\end{center}
%\end{figure}
%%%%%%%%%%%%%%%%%%%%%%%%%%%%%%%%%%%%%%%%%

%%%%%%%%%%%%%%%%%%%%%%%%%%%%%%%%%%%%%%%%%%%%%%%%%%%%%%%%%%%%%%%%%%%%%%%%%%%%%%%%%%%%%%%%%%%%%%%%%%%%%%%%%%%%%%%%%%%%%%%%%%%%%%%%%%%%%%%%%%%%%%%%%%%%%%%%%%%%%%%%%%%%%%%%%%%%%%%%%%%%%%%%%%%%%%%%%%%%%%%%%%

\section{Conclusion}

In order to measure the distribution of metastable states of spin glasses we have introduced two different algorithms. The first method employs traditional Monte Carlo techniques like flat-histogram, replica-exchange, and weighted histogram analysis. This method is able to find local minima over a wide range in energy, however, it has great difficulties at high energy and the statistical errors are comparatively large. For the second approach we designed an ensemble that allows a direct uniform sampling of metastable states. This method can efficiently be applied for the Edwards-Anderson model and yields very precise results.

We find that for the Sherrington-Kirkpatrick model our results are consistent with the analytical predictions. Unfortunately, statistical errors are substantial and our simulations were only able to access the whole energy range for small system sizes.

Investigating the Edwards-Anderson model by means of our novel method we were able to measure the distribution of metastable states with great precision. We found that the results -- suitably rescaled -- show very little dependence on the system size and we are, therefore, confident that also for much larger systems distributions very similar to the ones we describe would be obtained.

% \section*{Acknowledgements}


\begin{thebibliography}{1}%
\makeatletter
\providecommand \@ifxundefined [1]{%
 \@ifx{#1\undefined}
}%
\providecommand \@ifnum [1]{%
 \ifnum #1\expandafter \@firstoftwo
 \else \expandafter \@secondoftwo
 \fi
}%
\providecommand \@ifx [1]{%
 \ifx #1\expandafter \@firstoftwo
 \else \expandafter \@secondoftwo
 \fi
}%
\providecommand \natexlab [1]{#1}%
\providecommand \enquote  [1]{``#1''}%
\providecommand \bibnamefont  [1]{#1}%
\providecommand \bibfnamefont [1]{#1}%
\providecommand \citenamefont [1]{#1}%
\providecommand \href@noop [0]{\@secondoftwo}%
\providecommand \href [0]{\begingroup \@sanitize@url \@href}%
\providecommand \@href[1]{\@@startlink{#1}\@@href}%
\providecommand \@@href[1]{\endgroup#1\@@endlink}%
\providecommand \@sanitize@url [0]{\catcode `\\12\catcode `\$12\catcode
  `\&12\catcode `\#12\catcode `\^12\catcode `\_12\catcode `\%12\relax}%
\providecommand \@@startlink[1]{}%
\providecommand \@@endlink[0]{}%
\providecommand \url  [0]{\begingroup\@sanitize@url \@url }%
\providecommand \@url [1]{\endgroup\@href {#1}{\urlprefix }}%
\providecommand \urlprefix  [0]{URL }%
\providecommand \Eprint [0]{\href }%
\providecommand \doibase [0]{http://dx.doi.org/}%
\providecommand \selectlanguage [0]{\@gobble}%
\providecommand \bibinfo  [0]{\@secondoftwo}%
\providecommand \bibfield  [0]{\@secondoftwo}%
\providecommand \translation [1]{[#1]}%
\providecommand \BibitemOpen [0]{}%
\providecommand \bibitemStop [0]{}%
\providecommand \bibitemNoStop [0]{.\EOS\space}%
\providecommand \EOS [0]{\spacefactor3000\relax}%
\providecommand \BibitemShut  [1]{\csname bibitem#1\endcsname}%
\let\auto@bib@innerbib\@empty
%</preamble>
\bibitem [{Note1()}]{Note1}%
  \BibitemOpen
  \bibinfo {note} {Here, we have used the logarithm of the average of $N_S$ in
  order to be able to compare our result. However, since both $N_S$ and $\Omega
  (E)$ are expected to be log-normal distributed, in the rest of the paper we
  consider averages of logarithmic quantities.}\BibitemShut {Stop}%
\end{thebibliography}%


\begin{thebibliography}{50}

\bibitem{Bray_Moore}
A. J. Bray and M. A. Moore, J. Phys. C: Solid State Phys. {\bf 14}, 1313 (1981).

\bibitem{Li}
T. Li, Phys. Rev. B {\bf 24}, 6579 (1981).

\bibitem{Newman_Stein}
C. M. Newman and D. L. Stein, Phys. Rev. E {\bf 60}, 5244 (1999).

\bibitem{Burda_1}
Z. Burda, A. Krzywicki, O. C. Martin, and Z. Tabor, Phys. Rev. E {\bf 73}, 036110 (2006).

\bibitem{Burda_2}
Z. Burda, A. Krzywicki, and O. C. Martin, Phys. Rev. E {\bf 76}, 051107 (2007).

\bibitem{Waclaw_Burda}
B. Waclaw and Z. Burda, Phys. Rev. E {\bf 77}, 041114 (2008).

\bibitem{BaityJesi_Parisi}
M. Baity-Jesi and G. Parisi, Phys. Rev. B {\bf 91}, 134203 (2015).

\bibitem{Heuer}
A. Heuer, J. Phys. Condens. Matter {\bf 20}, 373101 (2008).

\bibitem{hoga}
S. Schnabel and W. Janke, Comput. Phys. Commun. {\bf 220}, 74 (2017).

\bibitem{SK}
D. Sherrington and S. Kirkpatrick, Phys. Rev. Lett. {\bf 35}, 1792 (1975).

\bibitem{EA}
S. F. Edwards and P. W. Anderson, J. Phys. F {\bf 5}, 965 (1975).

\bibitem{muca1}
B. A. Berg and T. Neuhaus, Phys. Lett. B  {\bf 267}, 249 (1991).

\bibitem{muca2}
B. A. Berg and T. Neuhaus, Phys. Rev. Lett. {\bf 68}, 9 (1992).

\bibitem{Wang_Landau}
F. Wang and D. P. Landau, Phys. Rev. Lett. {\bf 86}, 2050 (2001).

\bibitem{muca_wght_det}
W. Janke, in H. Fehske, R. Schneider, and A. Wei\ss e (Eds.), Lect. Notes Phys., Computational Many-Particle Physics, vol. 739 (Springer, Berlin, 2008), pp. 79-140.

\bibitem{parallel_temp1}
K. Hukushima and K. Nemoto, J. Phys. Soc. Jpn. {\bf 65}, 1604 (1996).

\bibitem{wham1}
A. M. Ferrenberg and R. H. Swendsen, Phys. Rev. Lett. {\bf 61}, 2635 (1988); Phys. Rev. Lett. {\bf 63}, 1195 (1989).

\bibitem{wham2}
S. Kumar, J. Rosenberg, D. Bouzida, R. H. Swendsen, and P. Kollman, J. Comput. Chem. {\bf 13}, 1011 (1992).

\bibitem{WTM}
J. Dall and P. Sibani, Comput. Phys. Commun. {\bf 141}, 260 (2001).

%\bibitem{parisi}
%G. Parisi, Phys. Rev. Lett. {\bf 43}, 1754 (1979).

%\bibitem{spin_chain_metastable}
%B. Derrida and E. Gardner, J. Physique {\bf 47}, 959 (1986).

%\bibitem{sim_anneal}
%S. Kirkpatrick, C. D. Gelatt, Jr., and M. P. Vecchi, Science 220 (1983) 671.

%\bibitem{ex_cl_app}
%A.K. Hartmann, Physica A 224 (1996) 480.

%\bibitem{basin_hopping}
%D. J. Wales and J. P. K. Doye, J. Phys. Chem. A 101 (1997) 5111.



\end{thebibliography}
\end{document}